\documentclass[sigconf]{acmart}
\AtBeginDocument{%
  }

\usepackage{balance}
\usepackage{amsmath}
\usepackage{threeparttable}
\usepackage{booktabs}
\usepackage{caption}
\usepackage{subcaption}
\usepackage{graphicx}
\usepackage{epstopdf}
\usepackage{stfloats}
\usepackage[most]{tcolorbox}
\usepackage{CJKutf8}
\usepackage{enumitem}
\usepackage{bbding}

\usepackage{algorithm,algorithmic}

\usepackage[utf8]{inputenc}
\usepackage{cleveref}
\crefname{section}{§}{§§}
\Crefname{section}{§}{§§}

\usepackage{pgfplots} 
\usepackage{scalefnt} 

\definecolor{color1}{RGB}{250,100,0}
\definecolor{color2}{RGB}{50,150,200}
\definecolor{color3}{RGB}{230,25,75}

\setcopyright{acmlicensed}
\copyrightyear{2025}
\acmYear{2025}
\setcopyright{cc}
\setcctype{by}
\acmConference[SIGIR '25]{Proceedings of the 48th International ACM SIGIR
Conference on Research and Development in Information Retrieval}{July 13--18, 2025}{Padua, Italy}
\acmBooktitle{Proceedings of the 48th International ACM SIGIR Conference on Research and Development in Information Retrieval (SIGIR '25), July 13--18, 2025, Padua, Italy}
\acmDOI{10.1145/3726302.3731951}
\acmISBN{979-8-4007-1592-1/2025/07}

\begin{document}

\title{Alleviating LLM-based Generative Retrieval Hallucination in Alipay Search} 

\author{Yedan Shen$^*$} 
\email{shenyedan.syd@antgroup.com}
\author[]{Kaixin Wu$^*\dagger$}
\thanks{$^*$Equal contribution. \quad $\dagger$Corresponding author.} 
\email{daniel.wkx@antgroup.com}
\affiliation{%
  \institution{Ant Group}
  \city{Hangzhou}
  \country{China}
}

\author{Yuechen Ding}
\email{dingyuechen.dyc@antgroup.com}
\author{Jingyuan Wen}
\email{wenjingyuan.wjy@antgroup.com}
\affiliation{%
  \institution{Ant Group}
  \city{Hangzhou}
  \country{China}
}



\author{Hong Liu}
\email{yizhou.lh@antgroup.com}
\author{Mingjie Zhong}
\email{mingjie.zmj@antgroup.com}
\affiliation{%
  \institution{Ant Group}
  \city{Hangzhou}
  \country{China}
}

\author{Zhouhan Lin}
\email{lin.zhouhan@gmail.com}
\affiliation{%
  \institution{LUMIA Lab, Shanghai Jiao Tong University}
  \city{Shanghai}
  \country{China}
}

\author{Linjian Mo}
\email{linyi01@antgroup.com}
\author{Jia Xu}
\email{steve.xuj@antgroup.com}
\affiliation{%
  \institution{Ant Group}
  \city{Hangzhou}
  \country{China}
}
\renewcommand{\shortauthors}{Yedan Shen et al.}

\begin{abstract}
Generative retrieval (GR) has revolutionized document retrieval with the advent of large language models (LLMs), and LLM-based GR is gradually being adopted by the industry.
Despite its remarkable advantages and potential, LLM-based GR suffers from hallucination and generates documents that are irrelevant to the query in some instances, severely challenging its credibility in practical applications.
We thereby propose an optimized GR framework designed to alleviate retrieval hallucination, which integrates knowledge distillation reasoning in model training and incorporate decision agent to further improve retrieval precision.
Specifically, we employ LLMs to assess and reason GR retrieved query-document (q-d) pairs, and then distill the reasoning data as transferred knowledge to the GR model. 
Moreover, we utilize a decision agent as post-processing to extend the GR retrieved documents through retrieval model and select the most relevant ones from multi perspectives as the final generative retrieval result. 
Extensive offline experiments on real-world datasets and online A/B tests on Fund Search and Insurance Search in Alipay demonstrate our framework's superiority and effectiveness in improving search quality and conversion gains.
\end{abstract}





\maketitle

\section{Introduction}
Document retrieval plays a vital role in large-scale search systems, potentially engage a user query \(q \in \phi\) to a series of relevant documents \(\{d_1, \cdots, d_k\} \in D\) from candidate documents $D$ \cite{ref1, ref22, ref23}. 
The traditional document retrieval approaches include sparse retrieval (SR) and dense retrieval (DR). SR focuses on building a compact inverted index with term matching metrics like TF-IDF \cite{ref6}, BM25 \cite{ref7}, or query likelihood \cite{ref12}. While DR trains dual-encoders to generate dense semantic embeddings for both query and documents \cite{ref3,ref4,ref8}, and subsequently retrieves the most relevant documents by embedding similarity. However, the traditional approaches are limited by the embedding space bottleneck and missing fine-grained interaction of query and document (q-d) pairs \cite{ref5, ref13}.  
Most recently, generative retrieval (GR) is an emerging paradigm for text retrieval, which employs a sequence-to-sequence encoder-decoder architecture to retrieve documents by directly generating their identifiers (DocIDs) \cite{ref9, ref10}. 
With the rise of large language models (LLMs) such as GPT \cite{ref19}, LLaMA \cite{ref18}, Qwen \cite{ref16, ref24, ref17}, LLM-based GR \cite{ref10,ref11,ref20} and other LLM-based IR applications \cite{ref-revelance1, ref-revelance2, ref-revelance3} have attracted significant research attention. By memorizing candidate documents within its model parameters \cite{ref1}, the LLM-based GR effectively leverages the capabilities of LLMs.
Fund Search and Insurance Search are two essential search scenarios in Alipay Search, aimed at providing users with relevant fund and insurance products based on their queries.
The user queries are often diverse, incorporating complex and nuanced intentions, which requires the retrieval model to possess a certain level of reasoning ability.
LLMs have demonstrated strong reasoning capabilities and the ability to handle complex tasks \cite{ref14,ref15,ref-kg,ref-test-time-compute}. Therefore, we aim to leverage LLM-based GR to better understand users' complex intentions and enhance the retrieval performance.

Despite LLM-based GR's remarkable advantages, it suffers from the inherent hallucination problem of LLMs, leading to the generation of irrelevant documents in retrieval.
To alleviate the hallucination, we introduce a novel optimized GR framework, which enhance generative retrieval in the following ways: a) \textbf{Knowledge Distillation Reasoning.} We first construct positive and negative retrieval instances (i.e., relevant and irrelevant q-d pairs) as the source data for reasoning. Next, we employ a larger LLM to generate explicit reasoning processes that explain why a document is relevant or irrelevant to the query for each q-d pair in the source data. These reasoning processes are then distilled into the smaller GR model as additional knowledge, enabling the GR model performance.
b) \textbf{Decision Agent.} For further generative retrieval precision enhancement, we introduce a decision agent as a fallback strategy. Specifically, we leverage traditional retrieval model to retrieve relevant documents for each GR retrieved document. Then, we leverage open-source LLMs to evaluate the relevance of these documents from multiple perspectives. Only the documents deemed relevant by the LLMs are retained as the final retrieval result. In summary, our contributions are as follows: 

\begin{itemize} [leftmargin=*]
    \item We propose an optimized framework for generative retrieval, which integrates knowledge distillation reasoning in model training and decision agent in post-processing to ease the hallucination phenomena in GR and improve the retrieval precision.
    \item Our framework requires no additional manually annotated data, is easily applicable to any existing GR models, and significantly enhances retrieval accuracy.
    \item Our approach has been successfully deployed online to power Fund Search and Insurance Search in Alipay, delivering substantial economic benefits.
\vspace{-1em}
\end{itemize}

\begin{figure*}[ht]
\raggedleft
\centerline{\includegraphics[width=18cm]{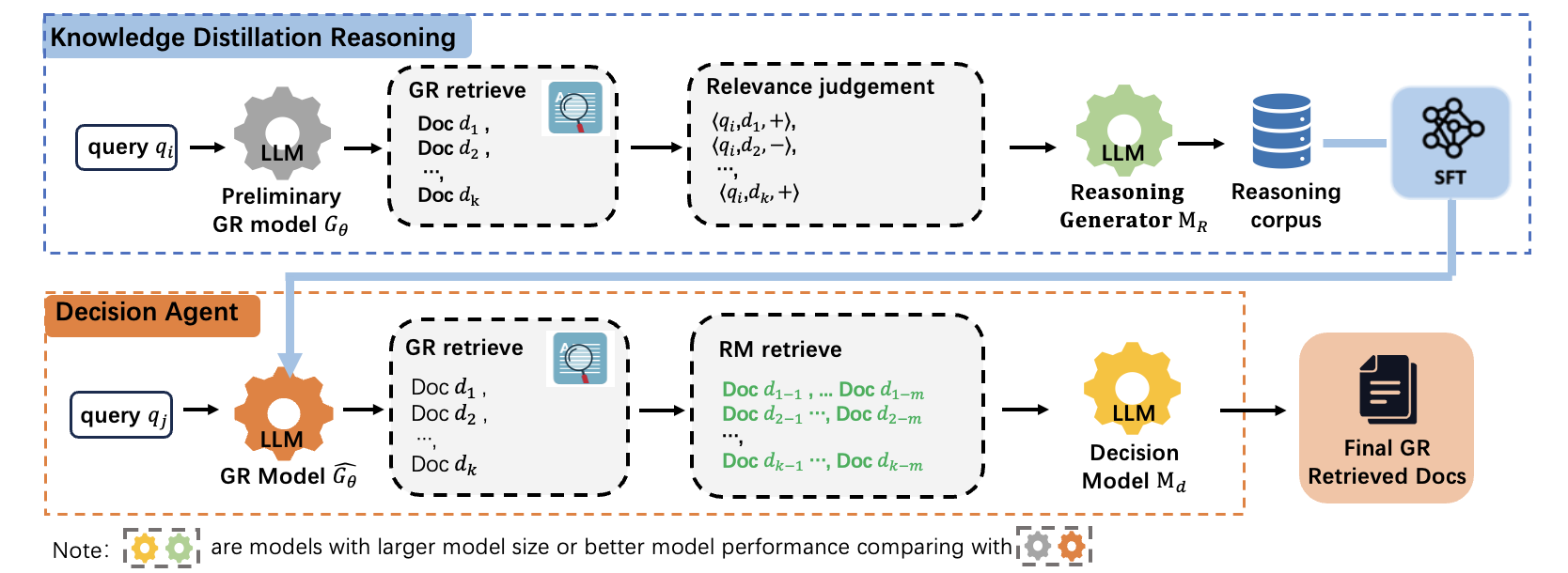}}
\vspace{-0.2cm}
\caption{An overview of our GR framework, which contains 2 modules: knowledge distillation reasoning and decision agent.}
\label{fig:GR_frame}
\vspace{-0.2cm}
\end{figure*}

\section{Methodology}
\subsection{Overview}
In this section, we first introduce the preliminary generative retrieval. Then we present the details of our optimized GR framework, including knowledge distillation reasoning and decision agent module as Figure~\ref{fig:GR_frame} shows. 

\subsection{Preliminary Generative Retrieval}
\label{sec:2.2}
The preliminary generative retrieval serves as the foundation of our framework. First, we construct training corpus \emph{T} with annotated relevant q-d pairs $\{\langle q,d^+ \rangle\}$ and structured knowledge $\{\langle d,k \rangle\}$ of the candidate documents \emph{D = $\{d_1, \cdots, d_n\}$}, n is the total number of candidate documents. Then we supervised fine-tune (SFT) the base LLM $M_\theta$ to preliminary GR model \emph{${G_\theta}$} with \emph{T}. The training loss is denoted as: 
\begin{equation}
    \emph{$\mathcal{L}_{GR}$} = - \sum_{i}{\sum_{l}^{L} y_i^l \log{p(\hat{y_i^l}|x_i,y_i^{<l})} }  \label{eqn-1},
\end{equation}
where $T=\{\langle q,d^+ \rangle, \langle d,k \rangle\}$, $(x_i,y_i) \in T$ is input-output pair, $L$ is the length of $y_i$, $y_i^l$ is the $l$-th token of $y_i$, $y_i^{<l}$ means the tokens before the $l$-th token in $y_i$, $\hat{y_i^l}$ is the predicted $l$-th token. 

\begin{algorithm}[!ht]
    \renewcommand{\algorithmicrequire}{\textbf{Input:}}
	\renewcommand{\algorithmicensure}{\textbf{Output:}}
	\caption{Knowledge Distillation Reasoning}
    \label{power}
    \begin{algorithmic}[1]
        \REQUIRE  Training corpus $T$, Query set $Q$, Base LLM for GR training $M_\theta$, Relevance judge model $\{M_1, \cdots, M_n\}$, Reasoning model $M_r$; 
	    \ENSURE GR model $\hat{G_\theta}$ trained by knowledge distillation reasoning; 

        \STATE Train preliminary GR model, $G_\theta \gets SFT(M_\theta, T)$ \COMMENT{\textcolor{blue}{\cref{sec:2.2}}}
        \STATE Sample reasoning source data $T_r = \{\langle q,d^+ \rangle\} \subseteq T$ 
        \FOR {each $q \in Q$}
            \STATE Retrieve $D = \{d_1, \cdots, d_k\} \gets G_\theta(q)$
            
            \STATE \parbox[t]{\dimexpr\linewidth-\algorithmicindent}{Relevance judgement ensemble 
            
            $D_{labeled} = \{\langle q,d_1,l_1 \rangle\, \cdots, \langle q,d_k,l_k \rangle\}, l_i \in \{+,-\} \gets Ensemble(M_1(q,D), \cdots, M_n(q,D))$}
            
            \FOR {each $\langle q,d_i,l_i \rangle\ \in D_{labeled}$}
                \IF {$l_i == -$}
                \STATE $T_r += \langle q,{d_i}^{-} \rangle $
                \ENDIF
            \ENDFOR
        \ENDFOR \COMMENT{\textcolor{blue}{\cref{sec:2.3.1}}}

        \FOR {each $\langle q,d_i^l \rangle \in T_r, l \in \{+,-\}$}
            \STATE Reasoning $r \gets M_r(\langle q,d_i^l \rangle)$
        \ENDFOR        

        \STATE \parbox[t]{\dimexpr\linewidth-\algorithmicindent}{Update training data 
        
        $T \rightarrow \hat{T} = T + \{\langle q,d^l, r \rangle\}, l \in \{+,-\}$       \COMMENT{\textcolor{blue}{\cref{sec:2.3.2}}}}

        \STATE Retrain GR model $\hat{G_\theta} \gets SFT(M_\theta, \hat{T})$ \COMMENT{\textcolor{blue}{\cref{sec:2.3.3}}}
        
        \STATE \textbf{return} $\hat{G_\theta}$.
    \end{algorithmic}
\label{algorithm_1}
\end{algorithm}

\subsection{Knowledge Distillation Reasoning}
Benefiting from the strong instruction-following capabilities of LLMs, we employ larger LLM to generate explicit reasoning data for smaller LLM-based GR model training. The detailed knowledge distillation reasoning algorithm is illustrated in Algorithm~\ref{algorithm_1}. 

\subsubsection{\textbf{Reasoning Source Data Construction.}}
\label{sec:2.3.1}
The reasoning source data consists of relevant and irrelevant q-d pairs. While the training corpus \emph{T} includes annotated relevant pairs $\{\langle q,d^+ \rangle\}$, our work mainly focuses on obtaining the irrelevant ones $\{\langle q,d^- \rangle\}$. 

As described in section ~\ref{sec:2.2}, we have trained a preliminary GR model ${G_\theta}$. Subsequently, we randomly sample a query set $Q$ from search logs, and utilize ${G_\theta}$ to generate retrieved documents $\{d_1, \cdots, d_k\}$ for each query $q$ in $Q$.
Aim to filter out irrelevant q-d pairs $\{\langle q,d^- \rangle\}$, we employ a series set of open-source LLMs $\{M_1, \cdots, M_n\}$ to identify the relevance for each {\emph{q}-$d_i$} pair, {$i \in 1,\cdots, k$} through Prompt 1. To ensure high-quality evaluation, only documents deemed irrelevant by all LLMs are classified as irrelevant. After relevance judgement, we construct the reasoning source data $R_s = \{\langle q,d^+ \rangle, \langle q,d^- \rangle \}$, where $\{\langle q,d^+ \rangle\}$ is sampled from training corpus \emph{T}.

\begin{tcolorbox}[colback=gray!10,
    colframe=black,
    width=8.5cm,
    arc=2mm, auto outer arc,
    boxsep={1pt},
    title={Prompt 1: Relevance Judgement of GR Retrieved Documents.},breakable,]			
    \textbf{Input:} \small In information retrieval scenario, the search query: <\emph{q}>, please identify if the retrieved document <$d_i$> is relevant to the query. 
    \vspace{0.1cm}
    
    <task-specific relevance judgement instruction>

    \vspace{0.1cm}
    \textbf{Output:} <Relevance judgement result.>
\end{tcolorbox}

\subsubsection{\textbf{Reasoning Generation.}}
\label{sec:2.3.2}
We employ a reasoning generator \emph{$M_r$} to generate high-quality reasoning process with Prompt 2. To ensure the quality of the reasoning data $R = {\{ \langle q,d^{+/-},r \rangle \}}$, the reasoning generator \emph{$M_r$} is a more powerful model (e.g., larger model size in the same series of LLM) comparing with the GR model \emph{${G_\theta}$}. 

\begin{tcolorbox}[colback=gray!10,
    colframe=black,
    width=8.5cm,
    arc=2mm, auto outer arc,
    boxsep={1pt},
    title={Prompt 2: Reasoning Generation.},breakable,]
    
    \textbf{Input:} \small In search scenario, it is known that the document <$d_i$> is <relevant/irrelevant> to the search query <\emph{q}>, please explain the reason.

    \vspace{0.1cm}
    \textbf{Output:} <Reasoning process $r$.>
\end{tcolorbox}

\subsubsection{\textbf{Distill Reasoning Data.}}
\label{sec:2.3.3}
With the reasoning data $R$, it becomes possible to enhance the model's understanding of the reasoning processes. In practice, we SFT GR model \emph{${\hat{G_\theta}}$} with the updated training dataset $\hat{T} = T + R$. 
The training loss of reasoning instruction is:
\begin{equation}
    \mathcal{L}_{reasoning} = - \sum_{i}{\sum_{l}^{L} r_i^l \log{p(\hat{r_i^l}|x_i,r_i^{<l})} },  \label{eqn-2} 
\end{equation}
where $(x_i,r_i) \in R$ is input-output pair, $x_i$ is the q-d pair with relevance judgement ${\langle q,d^{+/-}\rangle}$, $r_i$ is the retrieval rationale, $L$ is the length of $r_i$, $r_i^l$ is the $l$-th token of $r_i$, $r_i^{<l}$ means the tokens before the $l$-th token in $r_i$, $\hat{r_i^l}$ is the predicted $l$-th token. 

\subsection{Decision Agent}

Although we have enhanced the GR model training through knowledge distillation reasoning, the hallucination problem persists. Therefore, we introduce a decision agent in the post-processing stage for further precision improvement.

Given a GR retrieved document $d_i$ and the candidate documents $D$, we first leverage a retrieval model $RM$ to retrieve top-m relevant documents $\{d_{i-1}, \cdots, d_{i-m}\}$ for $d_i$. $RM$ can either be sparse or dense retrieval model. Then, we employ a powerful LLM $M_d$ to assess the relevance between query and the RM retrieved documents from multi perspectives with Prompt 3. In our scenario, structured information like product company, product type, and product duration are served as the various perspectives for decision model. Only the documents that are deemed relevant by all perspectives are retained as the final retrieval results.



\begin{tcolorbox}[colback=gray!10,
    colframe=black,
    width=8.5cm,
    arc=2mm, auto outer arc,
    boxsep={1pt},
    title={Prompt 3: Decision Prompt.},breakable,]			
    \textbf{Input:} \small Given the search query: <$q$>, the retrieved documents: $\{ \langle title: \$title, persp\_name: \$persp\_value\rangle, \cdots \}$, please output the titles of the relevant documents that satisfy the query's requirements from the perspective of <\emph{persp\_name}>. 

    \vspace{0.1cm}
    \textbf{Output:} <Relevant documents.>
\end{tcolorbox}

    

\section{Experiments}
\subsection{Experimental Setup}
\subsubsection{\textbf{Dataset and Metrics.}} Our training dataset consists of relevant q-d pairs and structured knowledge (e.g., product company, product type, product risk type), with approximately 250k records for fund retrieval and 200k for insurance retrieval. Additionally, we construct 20k reasoning data for each scenario and combine it with the original training corpus to train enhanced GR model.

For offline model validation, we sample high, medium, and low frequency queries from the search log as the test dataset. Experienced annotators label the relevance of the retrieved products to the respective queries, classifying each result as either relevant or irrelevant.
We use accuracy (ACC) as our offline evaluation metric.

\subsubsection{\textbf{Baseline Models.}} We compare our proposed GR model with the
following baselines:
\begin{itemize} [leftmargin=*]
    \item \textbf{BM25} is a representative sparse retrieval model that estimates the relevance based on term frequency, document length, and document frequency.
    \item \textbf{GR-baseline} is the preliminary Generative Retrieval model introduced in section ~\ref{sec:2.2}.
\end{itemize}

\subsubsection{\textbf{Implement Details.}} 
For preliminary GR model training, we take Qwen2.5-14B as GR backbone model and treat the product title as DocID. 
As for knowledge distillation reasoning, we employ Qwen2.5-72B and GPT-4o as the relevance judgement LMs. Additionally, Qwen2.5-72B is used to generate the reasoning processes. 
For the decision agent setup, we use BM25 as the retrieval model and Qwen2.5-32B as the decision model to generate the final GR model output.

\vspace{-0.5cm}
\begin{table}[t]
    \begin{subtable}{.5\linewidth}
      \centering
        \resizebox{!}{0.9cm}{
        \begin{tabular}{lcc}
            \toprule
                \textbf{Models} & \textbf{Fund} & \textbf{Insurance} \\
                \midrule
                  BM\_25 & 69.33\% & 47.33\%\\
                  GR-baseline & 83.83\% & 85.83\%\\
                \midrule
                  \textbf{Ours} & \textbf{87.17\%} & \textbf{90.05\%}\\
            \bottomrule
        \end{tabular}
        }
    \caption{Performance Comparison.}
    \label{table_1(a)}
    \end{subtable}%
    \begin{subtable}{.5\linewidth}
      \centering
        \resizebox{!}{0.9cm}{
        \begin{tabular}{lcc}
            \toprule
                \textbf{Models} & \textbf{Fund} & \textbf{Insurance} \\
                \midrule 
                \textbf{Ours} & \textbf{87.17\%} & \textbf{90.05\%} \\
                \midrule
                  w/o reasoning & 85.17\% & 87.66\%\\
                  w/o decision agent & 85.50\% & 86.17\%\\
            \bottomrule
        \end{tabular}
        }
    \caption{Ablation study.}
    \label{table_1(b)}
    \end{subtable} 
    \caption{Offline experiment result on ACC.}
    \vspace{-1cm}
\end{table}

\vspace{0.2cm}
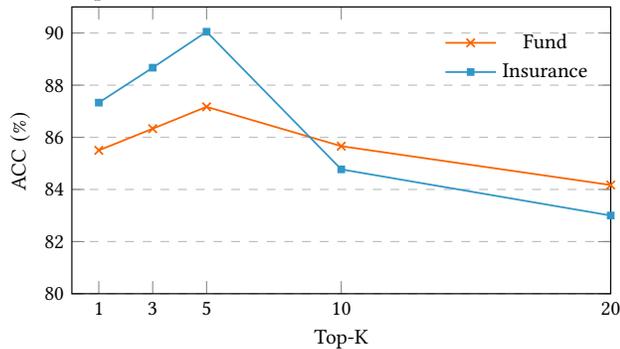
\begin{figure}[h] 
    \centering 
    \resizebox{1\columnwidth}{!}{  
        \begin{tikzpicture} 
        \scalefont{0.8} 
        \begin{axis}[
        sharp plot, 
        xmode=normal,
        xlabel=Top-K, 
        ylabel=ACC $(\%)$, 
        width=8cm, height=5cm,  
        xmin=0,xmax=20,  
        ymin=80, ymax=91,  
        xtick={1,3,5,10,20}, 
        ytick={80,82,84,86,88,90}, 
        xlabel near ticks, 
        ylabel near ticks, 
        ymajorgrids=true, 
        grid style=dashed, 
            legend style={at={(0.83,0.7)},anchor=south,fill=none,draw=none}, 
        ]
        
        \addplot+[semithick=2,mark=x,mark options={scale=1}, color=color1] plot coordinates { 
            (1,85.5)
            (3,86.33)
            (5,87.17)
            (10,85.66)
            (20,84.17)
        };
        \addlegendentry{Fund}
        
        \addplot+[semithick=2,mark options={scale=0.5}, color=color2] plot coordinates {
            (1,87.33)
            (3,88.67)
            (5,90.05)
            (10,84.77)
            (20,83)
        };
        \addlegendentry{Insurance} 
        \end{axis}
        \end{tikzpicture}
    }
    \setlength{\abovecaptionskip}{-0.3cm} 
    \caption{Comparison of Different Recall Numbers.} 
    \label{fig2}  
\vspace{-0.4cm}
\end{figure}

\subsection{Offline Experimental Results}
\subsubsection{\textbf{Offline Performance Comparison}}
From Table~\ref{table_1(a)}, it is evident that our method shows a significant improvement over the baseline GR model, with an absolute increase in ACC of 3.34\% on fund data and 4.22\% on insurance data, thereby validating the effectiveness of our approach. When compared to the retrieval baseline model BM25, the increase is even more substantial. Specifically, our method achieves a 17.84\% increase in ACC on fund data and a 42.72\% increase on insurance data, which highlights the necessity of incorporating GR models into fund and insurance search systems. Notably, insurance search queries tend to be more fuzzy, which explains the relatively poor performance of BM25 in this domain.

Moreover, for the decision agent, we analyze the impact of the number of input documents (Top-K) on ACC in Figure~\ref{fig2}. The results show that as Top-K increases, ACC first increase but then decreases. When Top-K becomes too large, it becomes challenging for the decision agent to effectively select the most relevant documents from the input set. Therefore, in our approach, we set Top-K to 5 to strike a balance between performance and input size.

\subsubsection{\textbf{Ablation Study}}
We conduct ablation experiments to analyze the contributions of knowledge distillation reasoning and the decision agent module. As illustrated by the experimental results in Table~\ref{table_1(b)}, removing the reasoning component leads to a 2\% drop in accuracy for fund retrieval and a 2.39\% decline for insurance retrieval. Additionally, the decision agent module demonstrates its effectiveness by contributing to a 1.67\% improvement in fund retrieval accuracy and a 3.88\% boost in insurance retrieval accuracy. These results highlight the indispensable roles of both components in our approach.



\subsection{Online A/B test}
\subsubsection{\textbf{Online Performance Comparison}}
To evaluate the efficacy of our framework, we conducted an online A/B test on the Fund Search and Insurance Search in Alipay.
Variant A represents the online baseline recall system, which combines both SR and DR recall paths. Meanwhile, Variant B incorporates our optimized GR framework as an additional recall path. 

Fund search includes exact and broad search: exact search matches fund names or codes, while broad search handles fuzzy queries, with the GR path affecting only the board search. In insurance search, as most queries are fuzzy, we evaluate only the overall performance.

Table~\ref{table_2} shows the performance improvement of our method in comparison to the previous online baseline. The result demonstrates statistically significant improvement in several key search metrics with a 95\% confidence level (p-value < 0.05), including Click Page View (Click\_PV), Click Unique Visitor (Click\_UV), Trade Count (Trade\_Count) and Trade Unique Visitor (Trade\_UV). 

\begin{table}[!ht]
\centering
\begin{threeparttable}
        \begin{tabular}{l|c|c|c} 
        \hline  
          \textbf{Metrics} & \multicolumn{2}{|c|}{\textbf{Fund Search}} & \textbf{Insurance Search} \\
          \hline
             & \textbf{Overall} & \textbf{Broad} & \textbf{Overall} \\
          \hline
          Click\_PV & +2.84\% & +12.71\% & - \\
          Click\_UV & +1.45\% & +11.06\% & - \\
          Trade\_Count & +1.31\% & +11.75\% & +1.89\%\\
          Trade\_UV & +0.99\% & +10.63\% & +2.07\%\\
        \hline
        \end{tabular}
\end{threeparttable}   
\caption{Online A/B Experiment Results.}\label{tab:tablenotes}
\vspace{-1cm}
\label{table_2}
\end{table}

\subsubsection{\textbf{Case Study}}
We present the generation cases from the Preliminary GR and our optimized GR framework, with factual errors highlighted in red. It can be observed that our model effectively reduces factual errors.

\begin{center}
\noindent\fbox{
    \parbox{1\linewidth}{
        \begin{CJK}{UTF8}{gbsn}
            {
            \small
                \textbf{Case 1}: \textbf{query}: 抗流感基金 {(Anti-flu fund)}
                
                \vspace{0.05cm}
                
                \textbf{GR-Baseline Output}: \textcolor{red}{国泰基金封闭}; 中银创新医疗混合C; 中银创新医疗混合A (\textcolor{red}{Guotai close-end fund}; BOC Innovative Healthcare Mixed C; BOC Innovative Healthcare Mixed A)  \begin{large} \textcolor{red}{\XSolidBrush} \end{large}
                
                \vspace{0.05cm}
                
                \textbf{Ours Output}: 中银创新医疗混合C; 中银创新医疗混合A; 华安医疗创新混合C (BOC Innovative Healthcare Mixed C; BOC Innovative Healthcare Mixed A; Huaan Healthcare Iinnovation Mixed C) \begin{large} \textcolor{red}{\CheckmarkBold} \end{large}

                \vspace{0.12cm}
                
                \hrule
                
                \vspace{0.12cm}

                \textbf{Case 2}: \textbf{query}: 单日意外保险 (One-day accident insurance)
                
                \vspace{0.05cm}
                
                \textbf{GR-Baseline Output}: 平安短期综合意外险; 运动意外无忧险; \textcolor{red}{1000万全年航空意外险} (Ping An Short-Term Comprehensive Accident Insurance; Sports Accident Worry-Free Insurance; \textcolor{red}{10 Million Annual Aviation Accident Insurance})  \begin{large} \textcolor{red}{\XSolidBrush} \end{large}
                
                \vspace{0.05cm}
                
                \textbf{Ours Output}: 平安短期综合意外险; 运动意外无忧险 (Ping An Short-Term Comprehensive Accident Insurance; Sports Accident Worry-Free Insurance) 
                \begin{large} \textcolor{red}{\CheckmarkBold} \end{large}
            }
        \end{CJK}
    }
}
\end{center}

\section{Conclusion}
In this paper, we propose an optimized GR framework to alleviate generative retrieval hallucination. Our approach consists of two key modules: knowledge distillation reasoning for GR model training and decision agent for post-processing. 
Specifically, we generate high-quality reasoning processes using a larger LM and distill the reasoning corpus into a smaller GR model to enhance its retrieval precision. Furthermore, we employ a decision agent to validate the relevance of GR retrieved documents. Together, these two mechanisms effectively filter out irrelevant documents, thereby significantly reducing retrieval hallucination.
Experimental results from both offline evaluations and online A/B tests conducted in two Alipay search scenarios show the effectiveness and scalability of our proposed method.


\bibliographystyle{ACM-Reference-Format}
\balance
\bibliography{sample-base}

\section*{COMPANY PORTRAIT}
Alipay is China’s leading digital interconnectivity and third-party payment open platform, with over 3 million merchants and institutions providing convenient services such as bill payments, transfers, online food ordering, online wealth management, and live shopping to 1 billion consumers.

\section*{PRESENTER BIOGRAPHY}
Presenter: Yedan Shen is an algorithm engineer at Alipay, focusing on research and development of large-scale search systems to improve user search experiences and boost search conversion gains. During her work in Alipay, she applied machine learning techniques to improve both retrieval and relevance in search systems.

\end{document}